\documentclass{pasj00}

\begin{document}
\SetRunningHead{Ota et al.}{Stellar Population of a $z=6.96$ LAE}
\Received{2010/03/25}
\Accepted{2010/06/17}

\title{\large Spitzer Space Telescope Constraint on the Stellar Mass of a $z=6.96$ Ly$\alpha$ Emitter}

\author{Kazuaki \textsc{Ota} %
}
\affil{Institute for Cosmic Ray Research, University of Tokyo, 5-1-5 Kashiwa-no-Ha, Kashiwa City, Chiba, 277-8582, Japan}
\email{ota@icrr.u-tokyo.ac.jp}

\author{Chun \textsc{Ly}, Matthew A. \textsc{Malkan}}
\affil{Department of Physics and Astronomy, Box 951547, UCLA, Los Angeles, CA 90095, USA}

\author{Kentaro \textsc{Motohara}}
\affil{Institute of Astronomy, University of Tokyo, 2-21-1 Osawa, Mitaka, Tokyo, 181-0015}

\author{Masao \textsc{Hayashi}}
\affil{National Astronomical Observatory of Japan, 2-21-1 Osawa, Mitaka, Tokyo, 181-8588}

\author{Kazuhiro \textsc{Shimasaku}}
\affil{Department of Astronomy, Graduate School of Science, University of Tokyo, 7-3-1 Hongo, Bunkyo-ku, Tokyo 113-0033}

\author{Tomoki \textsc{Morokuma\footnote{Research Fellow of the Japan Society for the Promotion of Science}}, Masanori \textsc{Iye}, Nobunari \textsc{Kashikawa}}
\affil{National Astronomical Observatory of Japan, 2-21-1 Osawa, Mitaka, Tokyo, 181-8588}



\and
\author{Takashi {\sc Hattori}}
\affil{Subaru Telescope, 650 North A'ohoku Place, Hilo, HI 96720, USA}


%


\maketitle

\begin{abstract}
We obtained mid-infrared 3.6 and 4.5 $\mu$m imaging of a $z=6.96$ Ly$\alpha$ emitter (LAE) IOK-1 discovered in the Subaru Deep Field, using {\it Spitzer Space Telescope} Infrared Array Camera observations. After removal of a nearby bright source, we find that IOK-1 is not significantly detected in any of these infrared bands to $m_{3.6\mu{\rm m}} \sim 24.00$ and $m_{4.5\mu{\rm m}} \sim 23.54$ at $3\sigma$. Fitting population synthesis models to the spectral energy distribution consisting of the upper limit fluxes of the optical to infrared non-detection images and fluxes in detection images, we constrain the stellar mass $M_*$ of IOK-1. This LAE could have either a mass as low as $M_* \lesssim 2$--$9 \times 10^{8} M_{\odot}$ for the young age ($\lesssim 10$ Myr) and the low dust reddening ($A_V \sim 0$) or a mass as large as $M_* \lesssim 1$--$4 \times 10^{10} M_{\odot}$ for either the old age ($> 100$ Myr) or the high dust reddening ($A_V \sim 1.5$). This would be within the range of masses of $z\sim 3$--6.6 LAEs studied to date, $\sim10^6$--$10^{10} M_{\odot}$. Hence, IOK-1 is not a particularly unique galaxy with extremely high mass or low mass but is similar to one of the LAEs seen at the later epochs.  
\end{abstract}

\section{Introduction}
Studies of stellar populations of galaxies at high-redshift frontiers such as Lyman break galaxies (LBGs) and Ly$\alpha$ emitters (LAEs) can constrain the scenarios of galaxy formation and evolution, by disentangling the histories of mass assembly and star-formation in the earliest stages of the universe.  Recent observations of rest-frame UV to optical continua of these galaxies at $z\sim3$--7 using deep optical to mid-infrared imaging from ground-based telescopes, {\it Hubble} and {\it Spitzer Space Telescope} suggest that both LBGs and LAEs consist of a range of types: from very young starbursts (a few to a few hundred Myr old, stellar mass of $\lesssim 10^8$--$10^9 M_{\odot}$) to massive evolved systems (a few hundred to $\sim1000$ Myr old, $\sim 10^{10}$--$10^{11} M_{\odot}$) with prominent Balmer/4000\AA~break that experienced much of their star-formation at earlier times \citep{Bradley08,Eyles05,Eyles07,Labbe06,Yan05,Yan06,Zheng09,SP05,Chary05,Finkelstein07,Finkelstein09,Gawiser06,Lai07,Lai08,Nilsson07,Pirzkal07,Ouchi09a}. The color-magnitude (corresponding to age-mass) distribution of LAEs tend to be very similar to the faint and blue end of LBGs. LAEs could be the lower mass extension of the LBG population \citep{Lai07,Lai08,Pirzkal07}. This implies that LAEs at the earlier cosmic epoch might evolve to be LBGs at later epochs. In any case, increasing our knowledge about both LBGs and LAEs and their relation is key to constraining the physical evolution of the entire galaxy population.

Statistics of stellar population studies of galaxies have been improved up to the redshift of $z\sim6$--7 or even higher. While such studies of $z\sim6$ galaxies are based on both some spectroscopically confirmed galaxies and many photometric candidates, all studies of $z\gtrsim7$ galaxies to date are still limited to photometric candidates. In this paper, we explore the stellar population of a spectroscopically confirmed $z = 6.96$ Ly$\alpha$ emitter, IOK-1 \citep{Iye06,Ota08}, which we discovered in the Subaru Deep Field (SDF; \cite{Kashikawa04}). We use {\it Spitzer} Infrared Array Camera (IRAC; \cite{Fazio04}) 3.6 and 4.5 $\mu$m imaging data of the SDF. This is the first study of the stellar population of a spectroscopically identified $z\simeq7$ galaxy. The paper is organized as follows. We describe the observations of the SDF and photometry in \textsection 2. Spectral energy distribution (SED) fitting of optical to mid-infrared data with population synthesis models is presented in \textsection 3. Finally, we discuss the results in \textsection 4 and conclude the study in \textsection 5. Throughout we adopt an $(\Omega_m, \Omega_{\Lambda}, h)=(0.3,0.7,0.7)$ concordance cosmology and AB magnitudes. 

\section{Observations and Photometry \label{OP}}
\subsection{Optical Imaging Data \label{Optical}}
For the photometry of optical (rest-frame UV) fluxes of IOK-1, we use deep broadband $BVRi'z'y$ images of SDF taken with Suprime-Cam \citep{Miyazaki02} on the Subaru Telescope. The $BV$ images are taken from the public SDF dataset version 1.0\footnote{Available from http://soaps.naoj.org/sdf/data/}. Meanwhile, we use $Ri'z'$ images that are even deeper than the public dataset and were recently obtained from 2005 to 2008 for variable object studies such as supernovae \citep{Poznanski07}, active galactic nuclei (Morokuma et al.~2010, in preparation), and high proper motion stars \citep{Richmond09}. Moreover, the $y$-band image of the SDF was obtained by \citet{Ouchi09b}. These images have seeing size of $\sim 0 \farcs 87$--$1 \farcs 13$. Source detection and photometry in each waveband were performed by using SExtractor \citep{BA96}. The aperture corrections to estimate the total $BVRi'z'$ magnitudes of a point source are obtained by running the SExtractor and comparing MAG$\_$AUTO's and MAG$\_$APER's of stellar objects in the images. They are $\sim 0.15$--$0.17$ mag. The $3\sigma$ limiting magnitudes measured with $2\arcsec$ apertures and with the aperture corrections applied are $(B, V, R, i', z')=(28.22, 27.54, 28.11, 27.35, 26.77)$ while the $y$-band image has the $3\sigma$ limiting magnitude of 26.4 measured by \citet{Ouchi09b} with $1.\arcsec8 $ aperture. The IOK-1 is not detected ($<2\sigma$ and not seen in the image by visual inspection) in any of the $BVRi'$ images due to the Ly$\alpha$ absorption by intergalactic medium (IGM) as in Figure 1. However, it is marginally detected and seen in the $z'$-band with $\sim3.2 \sigma$ significance (a total magnitude of $z'=26.70$) since the Ly$\alpha$ emission of IOK-1 is located at the very red edge of this waveband and because the $z'$-band image is 0.34 mag deeper than the public SDF $z'$ image used by \citet{Iye06} and \citet{Ota08}. In addition, the IOK-1 is detected in $y$-band at $7.4 \sigma$ significance (a total magnitude of $y=25.27$) and even selected as a $z'$-band dropout galaxy \citep[private communication]{Ouchi09b}.  

Also, imaging of IOK-1 with the narrowband NB973 filter (bandwidth 200\AA~centered at 9755\AA~corresponding to $z=6.9$--7.1 Ly$\alpha$ emission) was taken with Suprime-Cam in 2005. The photometry of NB973 was conducted in the same way as for the broadband images. The IOK-1 is detected at $6.5\sigma$ level (a total magnitude of NB973 $=24.4$) in this band \citep{Iye06,Ota08}. This total magnitude corresponds to the total flux of $F({\rm NB973}) \sim 4.0 \pm 0.5 \times 10^{-17}$ erg s$^{-1}$ cm$^{-2}$ by using
\begin{equation}
F({\rm NB973}) = f_{\nu}^{\rm NB} \frac{c}{\lambda_{\rm NB}^2} \Delta\lambda_{\rm NB}
\end{equation}
where $f_{\nu}^{\rm NB}$, $c$, $\Delta\lambda_{\rm NB}$ and $\lambda_{\rm NB}$ are the flux density in NB973, the speed of light, FWHM of NB973 filter (200\AA) and the central wavelength of NB973 filter (9755\AA), respectively. The uncertainty is a $1\sigma$ photometric error. 

\subsection{Optical Spectroscopy Data \label{Spec}}
The IOK-1 has a spectroscopically measured Ly$\alpha$ emission line flux of $F^{\rm spec}({\rm Ly}\alpha) \sim 2.0 \pm 0.4 \times 10^{-17}$ erg s$^{-1}$ cm$^{-2}$, where the uncertainty is a $1\sigma$ background noise in the spectrum \citep{Ota08}. However, this Ly$\alpha$ flux is under-estimated due to the slit loss of the flux at the time of spectroscopy of IOK-1. Thus, we estimate the slit loss as follows, assuming that Ly$\alpha$ and the UV continuum have the same morphology. First, we measure FWHM of IOK-1 in the NB973 image and that of the Ly$\alpha$ emission in the IOK-1 spectrum and confirm that they are similar ($\sim 1\arcsec$). On the other hand, a $0\farcs8 \times 14\farcs5$ slit was used at the time of the spectroscopy observation of IOK-1 \citep{Iye06,Ota08}. Hence, we compare the NB973 total flux of IOK-1 with the total flux measured in the NB973 IOK-1 image cut into the size of spectroscopy slit and obtain the fraction of the IOK-1 flux entering the slit, $\sim0.65$ or the slit loss of $\sim35$\%. Correcting the slit loss, we estimate the Ly$\alpha$ flux of IOK-1 to be $F^{\rm spec}_{\rm corr}({\rm Ly}\alpha) \sim 3.1 \pm 0.4 \times 10^{-17}$ erg s$^{-1}$ cm$^{-2}$.


\begin{table*}
\begin{center}
\caption{Ly$\alpha$ and Continuum Fluxes of IOK-1}
\begin{tabular}{lcc}
\hline
Estimation Method       & $F({\rm Ly}\alpha)$ & $f_{\nu}$(UV) \\    
& ($10^{-17}$ erg s$^{-1}$ cm$^{-2}$) & ($10^{-30}$ erg s$^{-1}$ cm$^{-2}$ Hz$^{-1}$)\\
\hline
spectrum and NB973 & $3.1\pm0.4$ & $1.7\pm1.6$\footnotemark[$*$] \\
spectrum and $z'$  & $3.1\pm0.4$ & $0.32\pm0.26$\footnotemark[$*$] \\
spectrum and $y$   & $3.1\pm0.4$ & $2.2\pm0.4$\footnotemark[$*$] \\
\hline
$z'$ and NB973     & $0.8^{+3.6}_{-0.8}$ & $5.6^{+7.2}_{-5.6}$\footnotemark[$\dagger$] \\
$y$ and NB973      & $2.0\pm1.0$ & $3.6\pm0.9$\footnotemark[$\dagger$] \\
\hline 
\multicolumn{3}{@{}l@{}}{\hbox to 0pt{\parbox{120mm}{\footnotesize
\par\noindent
\footnotemark[$*$] They are the continuum flux densities at all the wavelength within the passbands of the NB973, $z'$ and $y$ filters and correspond to the continuum magnitudes NB973(UV), $z'_{\rm cont}$ and $y_{\rm cont}$, respectively in \textsection \ref{LyaUV} and \ref{zycont} and Table 2. The $z'_{\rm cont}$ and $y_{\rm cont}$ are used for the SED-fitting of IOK-1 (See \textsection \ref{SED}).
\par\noindent
\footnotemark[$\dagger$] They are the continuum flux densities at the wavelength longward $z=6.96$ Ly$\alpha$ within the passband of the NB973.
}\hss}}
\end{tabular}
\end{center}
\end{table*}

\subsection{Ly$\alpha$ and UV Continuum Fluxes Estimated from Narrowband and Spectrum \label{LyaUV}}
If we assume that the NB973 total flux $F({\rm NB973})$ consists of Ly$\alpha$ and UV continuum fluxes, we can estimate UV continuum flux of IOK-1, $F({\rm UV})$, enclosed in the NB973 filter with
\begin{equation}  
F({\rm UV}) = F({\rm NB973}) - F^{\rm spec}_{\rm corr}({\rm Ly}\alpha). 
\end{equation}
This corresponds to the flux density of $f_{\nu}({\rm UV}) \sim 1.7 \pm 1.6 \times 10^{-30}$ erg s$^{-1}$ cm$^{-2}$ Hz$^{-1}$ as follows.
\begin{equation}\label{fnuUV}  
f_{\nu}({\rm UV}) = \frac{F({\rm UV})}{\Delta \lambda_{\rm UV}} \frac{\lambda_{{\rm Ly}\alpha}^2}{c}\\
\end{equation}
where $\Delta \lambda_{\rm UV} \sim 175$\AA~is the wavelength of UV continuum from the $z=6.96$ Ly$\alpha$ to the red edge of NB973 filter, and $\lambda_{{\rm Ly}\alpha}=(1+z)1216{\rm \AA} \sim 9680$\AA. This $f_{\nu}({\rm UV})$ corresponds to a UV continuum magnitude of NB973(UV) $ \sim 25.82_{-0.72}^{+3.08}$. Hence, $\sim 77_{-17}^{+21}$\% and $23^{+17}_{-21}$\% of the NB973 total flux come from Ly$\alpha$ emission and UV continuum, respectively.  

\subsection{Ly$\alpha$ and UV Continuum Fluxes Estimated from Narrowband and Broadband \label{LyaUV-NB-BB}}
Meanwhile, because we have detections of IOK-1 in both $z'$ and NB973 and know the exact redshift and the filter curves, we can also independently estimate the Ly$\alpha$ and UV continuum fluxes from the difference between the $z'$ and NB973 fluxes. We follow the same method as the equations (6) and (7) adopted by \citet{Taniguchi05} who estimated Ly$\alpha$ and UV fluxes of $z=6.6$ LAEs using $z'$ and narrowband filters. The UV continuum flux density of a $z=7$ LAE at the wavelength longward Ly$\alpha$ can be estimated by subtracting the total NB973 flux from the total $z'$ flux by using the following equation. 
\begin{equation}\label{fcont}
f_{\lambda}^{\rm image}({\rm UV}) = \frac{\Delta\lambda_{z'}f_{\lambda}^{z'} - 0.4 \Delta\lambda_{\rm NB}f_{\lambda}^{\rm NB}}{\Delta\lambda_{z'}^{\rm eff}}
\end{equation}
Here, $\Delta\lambda_{z'}=960$\AA~and $\Delta\lambda_{\rm NB}=200$\AA~are FWHM bandpasses of $z'$ and NB973 filters. The $f_{\lambda}^{z'}$ and $f_{\lambda}^{\rm NB}$ are the observed $z'$ and NB973 flux densities, respectively. The numerical factor of 0.4 is the relative transmittance of the NB973 filter with respect to the $z'$ filter. The $\Delta\lambda_{z'}^{\rm eff}$ is the wavelength from the red edge of the bandpass of NB973 filter to that of $z'$ filter. 

The IOK-1 has total magnitudes of $z'=26.70$ and NB973 $=24.4$, corresponding to the flux densities of $f_{\lambda}^{z'} \sim 2.4 \pm 0.8 \times 10^{-20}$ erg s$^{-1}$ cm$^{-2}$ ${\rm \AA}^{-1}$ and $f_{\lambda}^{\rm NB} \sim 2.0 \pm 0.3 \times 10^{-19}$ erg s$^{-1}$ cm$^{-2}$ ${\rm \AA}^{-1}$ where the uncertainties are $1\sigma$ photometric errors. Using the equation (\ref{fcont}), we obtain $f_{\lambda}^{\rm image}({\rm UV}) \sim 1.8^{+2.3}_{-1.8} \times 10^{-19}$ erg s$^{-1}$ cm$^{-2}$ ${\rm \AA}^{-1}$ or $f_{\nu}^{\rm image}({\rm UV}) = f_{\lambda}^{\rm image}({\rm UV}) \lambda_{{\rm Ly}\alpha}^2/c \sim 5.6^{+7.2}_{-5.6}\times 10^{-30}$ erg s$^{-1}$ cm$^{-2}$ Hz$^{-1}$.

Now the Ly$\alpha$ flux can be estimated by using
\begin{equation}\label{FLya}
F^{\rm image}({\rm Ly} \alpha ) = F({\rm NB973}) - f_{\lambda}^{\rm image}({\rm UV})\Delta \lambda_{\rm UV}
\end{equation}
where $\Delta \lambda_{\rm UV}$ is the same as the one in equation (\ref{fnuUV}), and the total UV continuum flux at the wavelength $\geq 1216$\AA~within the NB973 waveband is subtracted from the total NB973 flux. We obtain $F^{\rm image}({\rm Ly} \alpha) \sim 0.8^{+3.6}_{-0.8} \times 10^{-17}$ erg s$^{-1}$ cm$^{-2}$. The UV continuum flux density $f_{\nu}^{\rm image}({\rm UV})$ and the Ly$\alpha$ flux $F^{\rm image}({\rm Ly} \alpha)$ estimated from the difference between $z'$ and NB973 fluxes are consistent with $f_{\nu}({\rm UV})$ and $F^{\rm spec}_{\rm corr}({\rm Ly}\alpha)$ estimated in \textsection \ref{LyaUV} from the NB973 flux and slit-corrected Ly$\alpha$ flux from the spectrum. 

On the other hand, in the similar way, we also estimate the UV continuum flux density and Ly$\alpha$ flux from the difference between the NB973 and $y$-band fluxes, using the equations (\ref{fcont}) and (\ref{FLya}) with transmission curve of $y$-band instead of $z'$-band. In this case, we obtain $f_{\nu}^{\rm image}({\rm UV}) \sim 3.6 \pm 0.9 \times 10^{-30}$ erg s$^{-1}$ cm$^{-2}$ Hz$^{-1}$ and $F^{\rm image}({\rm Ly} \alpha) \sim 2.0 \pm 1.0 \times 10^{-17}$ erg s$^{-1}$ cm$^{-2}$. They are also consistent with $f_{\nu}({\rm UV})$ and $F^{\rm spec}_{\rm corr}({\rm Ly}\alpha)$ estimated in \textsection \ref{LyaUV} from the NB973 flux and slit-corrected Ly$\alpha$ flux from the spectrum.

\subsection{Continuum Fluxes in $z'$ and $y$ Bands \label{zycont}}
We can also estimate the continuum flux density of IOK-1 in $z'$-band $f_{\nu}(z'_{\rm cont})$ by subtracting the slit-corrected Ly$\alpha$ emission line flux $F^{\rm spec}_{\rm corr}({\rm Ly}\alpha)$ from the total $z'$ flux $F(z')$. Namely, 
\begin{equation}
f_{\nu}(z'_{\rm cont}) = \frac{F(z') - 0.4 F^{\rm spec}_{\rm corr}({\rm Ly}\alpha)} {\Delta\lambda_{z'}} \left( \frac{\lambda_{z'}^2}{c} \right)
\end{equation}
where $\lambda_{z'}=9190$\AA~is the central wavelength of the $z'$-band, and 0.4 is the relative transmittance of the NB973 filter with respect to the $z'$ filter. As in equation (\ref{fcont}), we have $F(z') = \Delta\lambda_{z'}f_{\lambda}^{z'} \sim 2.3 \pm 0.7 \times 10^{-17}$ erg s$^{-1}$ cm$^{-2}$ where the uncertainty comes from $1\sigma$ photometric error of $z'$-band image. Hence, we obtain $f_{\nu}(z'_{\rm cont}) \sim 3.2 \pm 2.6 \times 10^{-31}$ erg s$^{-1}$ cm$^{-2}$ Hz$^{-1}$ or correspondingly $z'_{\rm cont} \sim 27.63_{-0.63}^{+1.70}$ mag. In exactly the same way but using the $y$-band flux and transmission curve, we estimate the continuum flux density of IOK-1 in $y$-band to be $f_{\nu}(y_{\rm cont}) \sim 2.2 \pm 0.4 \times 10^{-30}$ erg s$^{-1}$ cm$^{-2}$ Hz$^{-1}$ or correspondingly $y_{\rm cont} \sim 25.57_{-0.20}^{+0.24}$ mag. All the measurements of the UV continuum flux density, Ly$\alpha$ flux and $z'$ and $y$-band continuum flux densities obtained in \textsection \ref{Spec}--\ref{zycont} are summarized in Table 1.

\subsection{Near-infrared Data \label{NIR}}

The near-infrared $J$-band (Ly et al.~2010, in preparation) and $K$-band \citep{Motohara08} images of the SDF were taken with the NOAO Extremely Wide-Field Infrared Imager (NEWFIRM; \cite{Probst08}) on the Kitt Peak National Observatory Mayall 4m Telescope and the Wide Field Camera on the United Kingdom Infrared Telescope (WFCAM; \cite{Henry03}), respectively. The FWHMs of the point spread function (PSF) of the images are estimated to be $2\arcsec$ and $1\farcs1$ in $J$ and $K$, respectively. Using the SExtractor, we estimate $3\sigma$ limiting magnitudes and aperture corrections for $J$-band ($3\arcsec$ aperture) and $K$-band ($2\arcsec$ aperture) in the same way as in \textsection \ref{Optical}. They are 22.75 and 0.24 mag in $J$ and 22.55 and 0.28 mag in $K$. Neither is IOK-1 detected in the $J$ nor $K$-bands ($<2\sigma$ and not seen in the images).
 

\begin{table*}
\begin{center}
\scriptsize
\caption{Photometry of IOK-1}
\begin{tabular}{lccccccccccc}\hline
Coordinate       & $B$ & $V$ & $R$ & $i'$ & $z'_{\rm cont}$ & NB973(UV) & $y_{\rm cont}$ & $J$ & $K$ & 3.6 $\mu$m & 4.5 $\mu$m \\ \hline    
13:24:18.4 +27:16:33 & $>$28.22 & $>$27.54 & $>$28.11 & $>$27.35 & $27.63_{-0.63}^{+1.70}$ & $25.82_{-0.72}^{+3.08}$ & $25.57_{-0.20}^{+0.24}$ & $>$22.51 & $>$22.27 & $>$23.99 & $>$23.54 \\ \hline 
\multicolumn{12}{@{}l@{}}{\hbox to 0pt{\parbox{185mm}{ 
Notes. Coordinates are in hours: minutes: seconds (RA) and degrees: arcminutes: arcseconds (DEC) J2000.0 equinox. The $BVRi'JK$, 3.6 $\mu$m and 4.5 $\mu$m fluxes are $3\sigma$ upper limits. The $z'_{\rm cont}$, NB973(UV) and $y_{\rm cont}$ are the continuum magnitudes calculated by subtracting the slit-corrected Ly$\alpha$ line flux measured in the spectrum of IOK-1 from each of the $z'$, NB973 and $y$ total fluxes (see \textsection \ref{LyaUV} and \ref{zycont} and Table 1). The $BVRi'K$ are $2''$ aperture magnitudes while $J$, 3.6 $\mu$m and 4.5 $\mu$m are $3''$ aperture magnitudes. Aperture corrections are applied to these magnitudes (see \textsection 2). All the broadband photometric measurements are used for the SED-fitting in \textsection 3 while NB973 flux is not since it is essentially the same as $y_{\rm cont}$ flux but its error is larger.}\hss}} 
\end{tabular}
\end{center}
\end{table*} 
 
\subsection{Mid-infrared Data \label{MIR}}

{\it Spitzer} IRAC imaging taken in 2006 in the 3.6, 4.5, 5.8 and 8.0$\mu$m wavebands covered most of the SDF field. The effective integration time was 1000 second per pixel for each band. The images were reduced using MOPEX software, and the final combined images have a pixel size of $0 \farcs 6$ pixel$^{-1}$, half their original size. The FWHM of the PSF of the images is estimated to be $\sim 2 \farcs 4$--$2 \farcs 6$. Using the SExtractor and in the same way as in \textsection \ref{Optical} and \textsection \ref{MIR}, we estimate the $3\arcsec$ aperture $3 \sigma$ limiting magnitude (and its corresponding aperture correction) of each combined image to be 24.54, 24.12, 22.48 and 22.34 (0.54, 0.58, 0.65 and 0.82 mag) in 3.6, 4.5, 5.8 and 8.0$\mu$m bands, respectively. IOK-1 is clearly not detected ($<2\sigma$ and not seen in the images) in any of 4.5, 5.8 and 8.0$\mu$m images.

In the 3.6$\mu$m image, we find a flux of $\sim4\sigma$ significance in a $3\arcsec$ aperture at the position corresponding to the coordinates of IOK-1. However, neither detection nor nondetection of IOK-1 in 3.6$\mu$m is immediately conclusive, because the object at the position could be either the flux of IOK-1, that of the extended tail of the nearby large stellar object, or a blend of the both. To check the reality of IOK-1 detection, we attempt to remove the neighboring stellar object by modeling and fitting its surface brightness profile with the GALFIT software \citep{Peng02}. 

To model the stellar object with GALFIT, we construct the PSF image in the 3.6$\mu$m band by stacking stellar objects randomly selected in the SDF image. We use the coordinates and magnitude of IOK-1 measured in the NB973 image as initial guess inputs required by GALFIT, since the NB973 image has higher spatial resolution. First, we run GALFIT on the NB973 image without fixing any input parameters, choosing a S$\acute{\rm e}$rsic profile. We then run GALFIT on the 3.6$\mu$m image with a S$\acute{\rm e}$rsic profile convolved with the PSF, adopting the parameters output by the fitting with NB973 as our initial guess for the input parameters and allowing them to vary. The neighboring stellar object is fitted and subtracted by GALFIT with this procedure. The resultant image shows that some flux is left but as weak as $\sim1.4\sigma$, consistent with non-detection ($<2\sigma$). Although the detection of IOK-1 is still difficult to judge, we conclude that the IOK-1 is not detected in 3.6$\mu$m. The photometry and images of IOK-1 in the optical to mid-infrared are summarized in Table 2 and Figure 1. The modeling and subtraction of the contaminating neighboring stellar source in the 3.6 $\mu$m image are shown in Figure 2.      

\begin{figure}
  \begin{center}
    \FigureFile(70mm,70mm){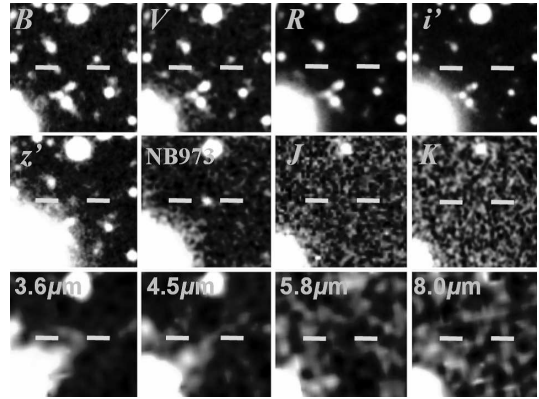}
  \end{center}
  \caption{The optical (Subaru/Suprime-Cam $BVRi'z'$ and NB973), near-infrared (Kitt Peak 4m/NEWFIRM $J$ and UKIRT/WFCAM $K$) and mid-infrared ({\it Spitzer}/IRAC 3.6, 4.5, 5.8 and 8.0 $\mu$m) images of IOK-1. Each image is $10\arcsec \times 10\arcsec$ on a side. The flux at the position of IOK-1 in the 3.6 $\mu$m image could be the blend of fluxes from both IOK-1 and extended tail of the neighboring large stellar source (See also Figure 2). See Figure 3 in \citet{Ouchi09b} for the y-band image of IOK-1.}\label{fig1}
\end{figure}

\begin{figure}
  \begin{center}
    \FigureFile(70mm,70mm){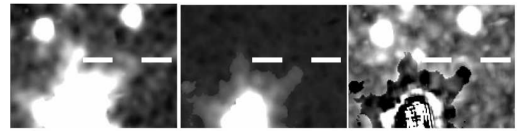}
  \end{center}
  \caption{Modeling and subtracting the contaminating stellar source near the position of IOK-1 in the 3.6 $\mu$m image, using GALFIT software \citep{Peng02}. ({\it Left}) The original image. ({\it Middle}) The surface brightness profile of the stellar source fitted and modeled by the GALFIT. ({\it Right}) The image after the subtraction was conducted.}\label{fig2}
\end{figure}

\section{SED Fitting to Population Synthesis Models \label{SED}}
The analysis of optical to mid-infrared images has shown that IOK-1 is not detected in these wavebands except for $z'$, NB973 and $y$ bands. Given the lack of any other detections at rest frame UV to optical wavelength, we cannot constrain important physical properties such as age and dust extinction. However, the upper limits on the IRAC data points can still provide a meaningful constraint, an upper limit on stellar mass. Since the IOK-1 is the only one galaxy spectroscopically identified to be at $z\simeq7$ at the present time, this constraint is particularly interesting, and we fit the SED of IOK-1 to stellar population synthesis models.

The SED of IOK-1 spanning rest-frame UV to optical wavelength is constructed from $3\sigma$ upper limits on the fluxes of Suprime-Cam $BVRi'$ bands, continuum fluxes around Ly$\alpha$, $z'_{\rm cont}$ and $y_{\rm cont}$, estimated from the $z'$ and $y$ fluxes by subtracting slit-corrected Ly$\alpha$ emission flux (See \textsection \ref{zycont} and Table 1), and $3\sigma$ upper limits on the fluxes of IRAC 3.6 and 4.5$\mu$m bands. This SED is used for fitting to stellar population synthesis models. The NB973(UV) flux is not used for the SED-fitting since it is essentially the same as $y_{\rm cont}$ flux but its photometric error is larger. We do not use 5.8 and 8.0$\mu$m data, either because they are not sufficiently deep. Meanwhile, template spectra to be compared are generated by the population synthesis models of \citet[hereafter, BC03]{BC03}. The Padova 1994 models preferred by BC03 are used. We assume a solar metallicity ($Z=Z{\odot}=0.02$) and a \citet{Salpeter55} initial mass function (IMF) with mass cutoffs of 0.1 and 100 $M_{\odot}$.

\begin{table*}
\begin{center}
\caption{Upper Limit on the Stellar Mass of IOK-1 from SED-fitting}
\begin{tabular}{ccccccccc}\hline
Reference\footnotemark[$*$] & $z_{\rm ref}$\footnotemark[$\dagger$] & $M_*^{\rm ref}$\footnotemark[$\ddagger$] & SFH &  Age  & $A_V$ & $\chi^2_{\nu}$ &	$M_*^{\rm IOK1}$\footnotemark[$\S$] & SFR\footnotemark[$\P$] \\	
           &	 & $(10^9M_{\odot})$ &  & (Myr) & (mag) &       & $(10^9M_{\odot})$ & $(M_{\odot}$yr$^{-1})$ \\ \hline
\multicolumn{9}{l}{Reasonable Model Parameters Typical of LAEs from Literatures} \\ \hline
L07  	   & 5.7 & 1.0--5.0 & SSP &   5 &  1.0  & 2.272 & $\leq 4.7_{-1.9}^{+1.2}$    & --- \\	
F09	       & 4.5 & 0.18     & SSP &  10 &  0.5  & 0.776 & $\leq 3.5_{-1.6}^{+2.8}$    &	--- \\ \hline	
L07	       & 5.7 & 4.5--6.9 & CSF &   5 &  1.5  & 2.837 & $\leq 11.4_{-4.7}^{+2.3}$   & $\leq 38.1_{-15.6}^{+7.6}$ \\	
G06, L08   & 3.1 & 0.3--0.5 & CSF & 100 &  0.0  & 1.733 & $\leq 1.9_{-0.94}^{+1.7}$	  & $\leq 21.9_{-10.8}^{+19.4}$ \\	
L07	       & 5.7 & 11       & CSF & 720 &  0.5  & 2.477 & $\leq 43.3_{-17.9}^{+11.1}$ & $\leq 19.2_{-7.9}^{+4.9}$ \\ \hline
\multicolumn{8}{l}{Variation in Age with $A_V$ Forced to be 0} \\ \hline
---	       & --- & --- & SSP &   5 &  0.0	& 1.058 & $\leq 0.36_{-0.18}^{+0.35}$ &	--- \\	
---	       & --- & --- & SSP &  10 &  0.0	& 1.121 & $\leq 0.88_{-0.44}^{+0.91}$ &	--- \\	
L07        & 5.7 & 5.1--14 & SSP & 100 & 0.0 & 8.824 & $\leq 22.1_{-10.4}^{+0.48}$ &	--- \\ \hline
---        & --- & --- & CSF &   5 &  0.0	& 1.423 & $\leq 0.23_{-0.13}^{+0.32}$ &	$\leq 46.9_{-25.4}^{+64.9}$ \\	
---        & --- & --- & CSF &  10 &  0.0  & 1.303 & $\leq 0.33_{-0.17}^{+0.39}$ &	$\leq 33.3_{-17.5}^{+39.2}$ \\
G06, L08   & 3.1 & 0.3--0.5 & CSF & 100 &  0.0  & 1.733 & $\leq 1.9_{-0.94}^{+1.7}$	  &	$\leq 21.9_{-10.8}^{+19.4}$ \\	
---        & --- & --- & CSF & 720 &  0.0  & 1.452 & $\leq 12.4_{-5.5}^{+13.0}$  &	$\leq 21.5_{-9.5}^{+22.5}$  \\ \hline
\multicolumn{9}{@{}l@{}}{\hbox to 0pt{\parbox{140mm}{\footnotesize
\par\noindent
\footnotemark[$*$] The literatures used to obtain parameters of the BC03 model typical of LAEs at $z=3.1$--5.7: L07 \citep{Lai07}, F09 \citep{Finkelstein09}, G06 \citep{Gawiser06} and L08 \citep{Lai08}.
\par\noindent
\footnotemark[$\dagger$] Redshifts of the LAE samples in the literatures.
\par\noindent
\footnotemark[$\ddagger$] Typical stellar mass of the LAEs in the literatures.
\par\noindent
\footnotemark[$\S$] Upper limit on stellar mass of IOK-1 with errors corresponding to 68\% confidence intervals with $\Delta \chi^2=1$.
\par\noindent
\footnotemark[$\P$] Upper limit on SFR of IOK-1 at $z=6.96$ and its 68\% confidence interval obtained from the age and $M_*^{\rm IOK1}$. If $A_V>0$, the dust extinction is applied to the SFR using \citet{Calzetti00} law.
}\hss}}
\end{tabular}
\end{center}
\end{table*}

In the SED-fitting, we first choose reasonable values for star formation history (SFH), age and dust reddening ($A_V$), which can represent a LAE's physical properties, except stellar mass and find the values of the mass which go through the $z'_{\rm cont}$ and $y_{\rm cont}$ and do not violate either IRAC upper limits. We take the largest value of the mass the SED-fitting returns as the upper limit on the stellar mass. We then try a few variations in age to examine the change in the mass upper limit. We refer to the results of SED-fitting of LAEs at $z=3.1$, 4.5 and 5.7 performed by \citet{Gawiser06}, \citet{Lai08}, \citet{Finkelstein09} and \citet{Lai07} as the reasonable values for model parameters of our SED-fitting. They all used BC03 model with SFH of simple stellar population (SSP) or constant star formation rate (CSF), Salpeter IMF, solar metallicity and \citet{Calzetti00} dust extinction law for all or part of their studies, which we adopt for our study. 

First, \citet{Gawiser06} performed SED-fitting of stacked flux of 18 spectroscopically confirmed LAEs at $z=3.1$, whose rest frame UV to optical continua are too dim to detect, and obtained an average age of $\sim 90$ Myr, $A_V=0.0_{-0.0}^{+0.1}$ and stellar mass of $M_* \sim 5 \times 10^8 M_{\odot}$ for the CSF model. Similarly, fitting the stacking of 52 $z=3.1$ LAEs, undetected in IRAC (rest frame near-infrared), to the CSF models, \citet{Lai08} found an age of 160 Myr, $A_V \sim 0$ and $M_* \sim 3 \times 10^8 M_{\odot}$. Hence, we choose an age of 100 Myr and $A_V=0$ for a CSF model as a set of reasonable parameters typical of $z=3.1$ LAEs. 

Meanwhile, \citet{Finkelstein09} studied stellar population of 14 $z=4.5$ LAEs, 5 out of which have the best-fit results with $Z=Z_{\odot}$ SSP models. Four of them have ages of 2.5--15 Myr (9 Myr on average), $A_{1200}\sim 0.8$--3.0 (equivalently $A_V\sim 0.3$--1.0; 0.58 on average) and $M_*\sim 0.84$--$2.9 \times 10^8 M_{\odot}$ ($1.8 \times 10^8 M_{\odot}$ on average), while one is 40 Myr old with $A_V \sim 1.4$ and $M_* \sim 4.8 \times 10^9 M_{\odot}$. \citet{Finkelstein09} also performed Monte Carlo simulation to obtain more accurate SED-fitting results as the most likely models. In the case of $Z=Z_{\odot}$ SSP, the ages, $A_V$'s and masses of the most likely models are similar to those of the best-fit models. Thus, we take an age of 10 Myr and $A_V=0.5$ for an SSP model as a set of reasonable parameters typical of $z=4.5$ LAEs. 

Finally, \citet{Lai07} fitted the SEDs of three $z=5.7$ LAEs to both SSP and CSF models. For $Z=Z_{\odot}$, they obtained age $=$ (4.8, 3.2, 4.4) Myr, $A_V=(1.1$, 1.7, 1.1) and  $M_*=$ (2.4, 5.0, 1.0) $\times 10^9 M_{\odot}$ for SSP and age $=$ (5.0, 4.8, 720) Myr, $A_V=(1.5$, 1.7, 0.4) and  $M_*=$ (4.5, 6.9, 11) $\times 10^9 M_{\odot}$ for CSF, where we convert $E(B-V)$ they obtained into $A_V$ using \citet{Calzetti00} dust extinction law. Therefore, we adopt an age of 5 Myr and $A_V=1.0$ for SSP model and an age of 5 Myr and $A_V=1.5$ for CSF model as a set of reasonable parameters typical of $z=5.7$ LAEs. We also try the older age scenario of 720 Myr and $A_V=0.5$ for CSF for completeness. 

All the reasonable parameters (SFH, age and $A_V$) chosen for our SED-fitting as well as stellar mass, which are typical of LAEs at $z=3.1$, 4.5 and 5.7, are shown in Table 3. In the SED-fitting, we regard these parameters as fixed input assumptions of the models. Then, we compare all the SED-fitting results and investigate how the upper limit on the stellar mass of IOK-1 changes. To perform SED-fitting to the models, we use the public photometric redshift code of {\it Hyperz} v1.1 \citep{Bolzonella00}. The IGM absorption blueward Ly$\alpha$ is applied to each BC03 model using the prescription of \citet{Madau95}. When the effect of dust reddening $A_{V}$ is taken into account, the \citet{Calzetti00} dust obscuration formula is used. Fixing the redshift to be $z=6.96$ and age and $A_V$ to be the reasonable values in Table 3, we minimize the reduced $\chi^2_{\nu} = \chi^2/\nu$ with
\begin{equation}\label{chi2}
\chi^2 (z) = \sum_{i=1}^{N_{\rm filters}} \left[ \frac{F_{{\rm obs},i} - b \times F_{{\rm temp},i}(z)}{\sigma_i} \right]^2
\end{equation}
where the $\nu$, $F_{{\rm obs},i}$, $F_{{\rm temp},i}$, $\sigma_i$ and $b$ are the number of degrees of freedom, the observed and BC03 template fluxes and their uncertainty in filter $i$ and a normalization factor. 

We conduct the SED-fitting in the following way. First, as the inputs for {\it Hyperz}, we treat the upper limits as the fluxes having the $3\sigma$ values and errors of $\pm 1 \sigma$. When calculating $\chi^2$, {\it Hyperz} considers that $F_{{\rm obs},i}=3\sigma$ and $\sigma_i=1\sigma$ for the upper limits for the equation (\ref{chi2}) as if these upper limits were detection points. When the {\it Hyperz} returns the SED-fitting results, some of them violate either or both of IRAC upper limits ({\it i.e.}, $b \times F_{{\rm temp}} >$ IRAC $3\sigma$ flux). Thus, if the SED-fitting violates either IRAC upper limits, we penalize the $\chi^2$ ({\it i.e.}, increase the $\chi^2$) by reducing the normalization factor $b$ by the amount that makes the $b \times F_{{\rm temp}}$'s at $3.6\mu$m and $4.5\mu$m become the flux values right below the IRAC $3\sigma$ limits.

Meanwhile, when calculating $\chi^2/\nu$, {\it Hyperz} considers that $\nu =$ the number of filters $- 1$, including the nondetection data points ({\it i.e.}, upper limits) in the equation (\ref{chi2}), where the 1 corresponds to the number of interesting parameter ({\it i.e.}, the normalization factor $b$). However, because we have two detection data points ($z'$ and $y$) and try to fit one parameter (stellar mass or the normalization factor $b$) by fixing all the other parameters, we regard the number of degrees of freedom as $\nu = 1$ and calculate $\chi^2/\nu$ by ourselves, removing the nondetection data from the equation (\ref{chi2}). The SED-fitting results are shown in Table 3 and Figure 3.


\begin{figure*}
  \begin{center}
    \FigureFile(154mm,154mm){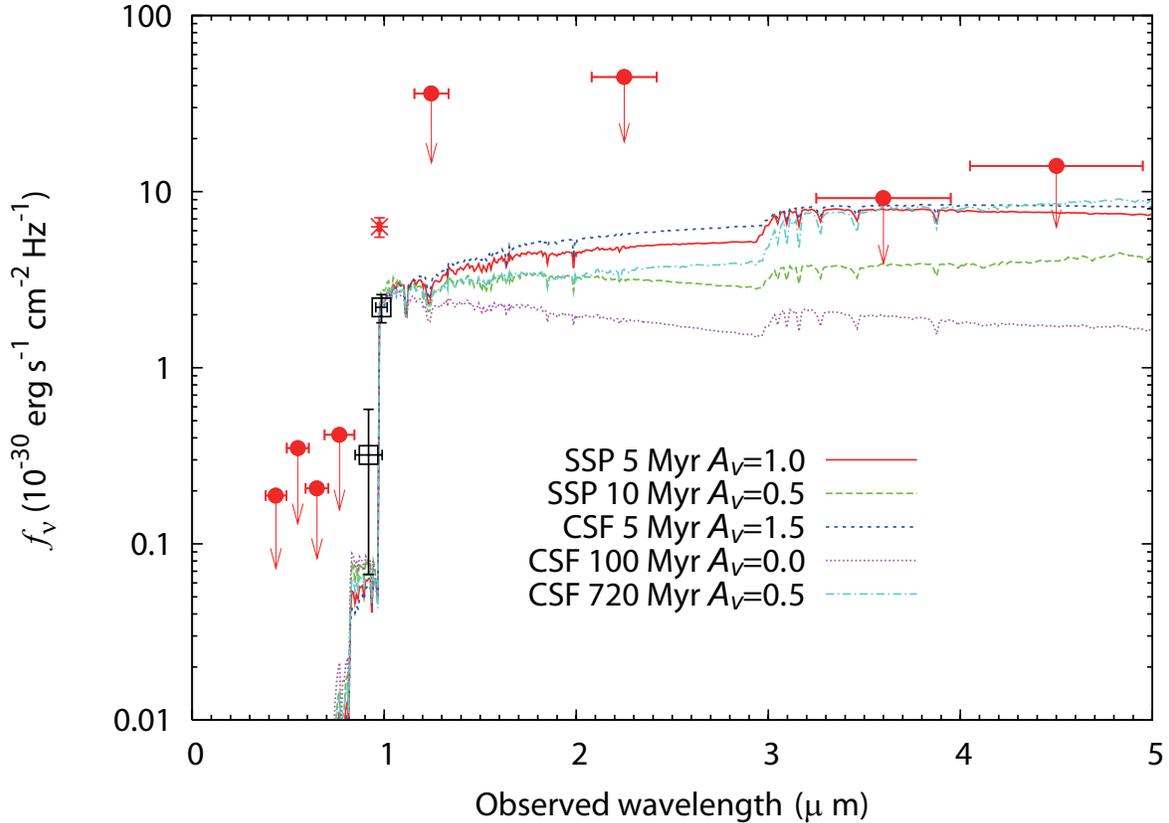}
  \end{center}
  \caption{The results of SED-fitting using the reasonable parameters (SFH, age and $A_V$) chosen in \textsection 3 and shown in Table 3. The lines are the fitted BC03 models while the filled circles are the photometric measurements in $BVRi'JK$, 3.6 $\mu$m and 4.5 $\mu$m. The measured fluxes in the detection bands $z'_{\rm cont}$ and $y_{\rm cont}$ are shown with the blank squares. The fluxes with arrows are the $3\sigma$ upper limits. The fluxes in $z'_{\rm cont}$ and $y_{\rm cont}$ and their vertical error bars are the continuum fluxes estimated by subtracting the Ly$\alpha$ flux and associated photometric errors (See \textsection \ref{zycont}). The asterisk denotes the NB973 total flux including the Ly$\alpha$ emission. The horizontal error bars indicate the bandwidth of the filters.}\label{fig3}
\end{figure*}

\section{Discussion}
\subsection{Constraint on Stellar Mass \label{age and mass}}
We estimate the upper limit on stellar mass $M_{*}$ of IOK-1 from the SED-fitting including the upper limits on the IRAC 3.6 and 4.5 $\mu$m bands. We use the normalization factor $b$ of each BC03 model in the SED-fitting returned by $Hyperz$ (see equation [\ref{chi2}]). The stellar mass is calculated with 
\begin{equation}\label{mass}
M_{*}=b \times \frac{(2 \times 10^{-17}) 4 \pi D_{L}^2}{L_{\odot}} \times M_{*}^{\rm model}
\end{equation}
where $D_{L}$ is the luminosity distance, $M_{*}^{\rm model}$ is the stellar mass of the template spectrum given by the BC03 code, and $L_{\odot}=3.826\times 10^{33}$ erg s$^{-1}$ is the solar luminosity (M. Bolzonella 2009, private communication; \cite{LS09,SLA09}). All the stellar masses obtained from the reasonable parameters for SFH, age and $A_V$ are shown in Table 3.

Also, for the CSF models, we can estimate the upper limit on the dust extinct star formation rate (SFR) of IOK-1 at $z=6.96$ from the SED-fitting and compare it with the observed SFR to exclude the BC03 models that give inconsistent SFRs, if any. We calculate intrinsic SFR of IOK-1 from the ages $t$, the total masses $M_{\rm tot}$ and the relations $M_{\rm tot} = {\rm SFR} \times t$, and then apply \citet{Calzetti00} dust extinction law with $A_V$ to obtain the dust extinct SFR. The $M_{\rm tot}$ is the sum of the mass currently in stars, in stellar remnants, and in gas returned to the interstellar medium by evolved stars, while the stellar mass $M_*$ is the mass currently in stars. The total mass is determined by the BC03 model, depending on the choise of SFH. The total mass of IOK-1, $M_{\rm tot}$, is obtained by substituting the total mass $M_{\rm tot}^{\rm model}$ of each template spectrum given by the BC03 code for the $M_*^{\rm model}$ in the equation (\ref{mass}). Since the $M_{\rm tot}$ is an upper limit on the total mass, the corresponding SFR is also an upper limit. The SFRs are also shown in Table 3. Note that because we fix the age and $A_V$ for the SED-fitting, the SFR is automatically determined, not changing the number of degrees of freedom, right after the $M_{\rm tot}$ (or the normalization $b$, which is only one free parameter in the SED-fitting) is obtained. 

Meanwhile, the observed SFR estimated from Ly$\alpha$ emission flux $F({\rm Ly}\alpha)$ measured in the spectrum of IOK-1 is SFR(Ly$\alpha) \sim 10 M_{\odot}$ yr$^{-1}$, while the SFR calculated from the UV continuum flux $F({\rm UV})=F({\rm NB973})-F({\rm Ly}\alpha)$ is SFR(UV$) \sim 36 M_{\odot}$ yr$^{-1}$ \citep{Ota08}. These SFRs are not corrected for the slit loss, 0.35, estimated in \textsection \ref{Spec}. If we instead use the fluxes corrected for the slit loss, $F({\rm Ly}\alpha)/(1-0.35)$ and $F({\rm NB973})-F({\rm Ly}\alpha)/(1-0.35)$, the SFRs become SFR(Ly$\alpha) \sim 15.8 M_{\odot}$ yr$^{-1}$ and SFR(UV$) \sim 16.5 M_{\odot}$ yr$^{-1}$. The SFR(Ly$\alpha$) is a lower limit, since the effects of dust extinction and IGM absorption of the line are not corrected. Also, the SFR(UV) is not corrected for dust extinction, and its intrinsic value could be larger if IOK-1 has dust. All the upper limits on SFR obtained from the SED-fitting are consistent with the observed SFR(UV) as seen in Table 3.

Figure 3 shows that the fitted BC03 template spectra go through $z'$ and $y$ fluxes around the Lyman break and do not violate either IRAC upper limits for all the models considered. As an overall trend common to both SSP and CSF seen in Table 3, the larger $A_V$ results in the larger mass. Table 3 also shows the typical stellar masses of $z=3.1$--5.7 LAEs obtained from the literatures. The upper limit on IOK-1 mass is slightly larger than masses of $z=3.1$--5.7 LAEs, and it can be a similar system to $z=3.1$--5.7 LAEs if the IOK-1 has similar SFH, age and $A_V$. The IOK-1 could have a stellar mass of $M_* \lesssim 10^9 M_{\odot}$ for SSP and a wider range of masses $M_* \lesssim 10^{9-10} M_{\odot}$ for CSF due to the wider range of reasonable values for ages and $A_V$ typical of $z=3.1$--5.7 LAEs. The largest mass is $M_* \lesssim 4.7_{-1.9}^{+1.2} \times 10^9 M_{\odot}$ for 5 Myr $A_V=1.0$ SSP model and $M_* \lesssim 4.3_{-1.8}^{+1.1} \times 10^{10} M_{\odot}$ for 720 Myr $A_V=0.5$ CSF model. Both of them corresponds to the young low mass (SSP) and old massive (CSF) scenarios for $z=5.7$ LAEs obtained by \citet{Lai07}.

\subsection{Change in Mass with Variation in SFH and Age}
We have chosen parameters, typical of $z=3.1$--5.7 LAEs, for SFHs, ages and $A_V$ obtained from the previous studies to see what the stellar mass of IOK-1 could be if we assume that IOK-1 has similar SFHs, ages and $A_V$. We have confirmed that the higher $A_V$ apparently results in larger mass. However, effects of different SFHs and ages in mass are not as obvious as the case of $A_V$. Hence, to examine how the upper limit on the stellar mass changes with SFH and age more accurately, we now perform the SED-fitting with the variation in age (5, 10, 100 and 720 Myr) for SSP and CSF while fixing $A_V$ to be zero. Table 3 also shows the results of this SED-fitting. We find that the fitted BC03 template spectra do not violate either IRAC upper limits and go through $z'$ and $y$ fluxes around the Lyman break for all the models considered except for the 720 Myr SSP model. We also try ages of 100--720 Myr for SSP and find that the SSP model spectra with $>100$ Myr ages violate IRAC upper limits if they go through $z'$ and $y$ fluxes and vice versa. Hence, we exclude the 720 Myr SSP model from the further discussion. 

Table 3 suggests that the SSP models give larger mass than CSF. The difference in mass between SSP and CSF becomes larger as age gets older (factors of $\sim1.6$, 2.7 and 12 for 5, 10 and 100 Myr, respectively). Also, as naturally expected and seen in Table 3, the older ages result in the larger masses. For ages $\leq 10$ Myr, both SSP and CSF give $M* \lesssim 10^8 M_{\odot}$ while the masses are $M* \lesssim 10^{9-10} M_{\odot}$ for ages $\geq 10$ Myr. The largest masses are $M* \lesssim 2.2_{-1.0}^{+0.05} \times 10^{10} M_{\odot}$ for SSP and $M* \lesssim 1.2_{-0.6}^{+1.3} \times 10^{10} M_{\odot}$ for CSF. In addition, the limit on SFR is larger for the younger age because the mass should be assembled in the shorter time. All the upper limit SFRs inferred from the CSF models are consistent with the observed SFR of IOK-1, SFR(UV$) \sim 16.5 M_{\odot}$ yr$^{-1}$. 

Eventually, according to Table 3, IOK-1 can have either a mass limit as low as $M_* \lesssim 2$--$9 \times 10^{8} M_{\odot}$ for the young age ($\sim 5$--10 Myr) and the low dust reddening ($A_V \sim 0$) or a mass limit as large as $M_* \lesssim 1$--$4 \times 10^{10} M_{\odot}$ for either the old age ($>100$ Myr) or the high dust reddening ($A_V \sim 1.5$). The previous studies have shown that the LAE populations at $z\sim3$--6.6 are very diverse including the LAEs from young, low mass, low dust ones (a few Myr old, $\sim 10^6$--$10^8 M_{\odot}$, $A_V\sim0$--0.5) to old, massive, dusty ones (a few hundred Myr old, $\sim 10^9$--$10^{10} M_{\odot}$, $A_V\gtrsim1$) \citep{SP05,Chary05,Finkelstein07,Finkelstein09,Gawiser06,Lai07,Lai08,Nilsson07,Pirzkal07,Ouchi09a}. For example, among these LAEs, ones at the highest redshift close to $z=7$ studied to date are $z\sim6.6$ LAEs, HCM 6A \citep{HCM02,Chary05} and Himiko \citep{Ouchi09a}. The HCM 6A was found to have a stellar mass of $\sim 8.4 \times 10^8$--$6.8 \times 10^9 M_{\odot}$ while the Himiko has $\sim 0.9$--$5.0 \times 10^{10} M_{\odot}$. The IOK-1 could be either a galaxy similar to the HCM 6A or the Himiko. Consequently, IOK-1 is not a particularly unique galaxy with extremely high mass or low mass, but its mass could be within the typical range of masses expected for LAEs.

\subsection{Impact of Nebular Emissions on the Mass}
Our discussions so far have been based on the SED-fitting without considering an impact of the nebular emissions. However, if we include fluxes from the nebular emissions in the population synthesis models, the stellar mass limit of IOK-1 would be lower. According to \citet{SdB09}, who also used BC03 models assuming \citet{Salpeter55} IMF and solar metallicity, the stellar masses of $z\sim6$ galaxies studied by \citet{Eyles07} would be on average $\sim 30$\% lower if they include nebular emissions in the model SEDs. In addition, although they used different stellar population models with lower metallicities (and also a slightly different cosmology from WMAP3), \citet{ZBL08} estimated the ratio of net nebula emission flux to stellar emission flux, $f_{\rm neb}/f_{\rm stars}$ to be $\sim 0.3$ for a $z=7$ starburst galaxy ($\sim 0.3$--0.5 for $z=6$) in {\it Spitzer}/IRAC 3.6 and 4.5 $\mu$m bands. This means the mass of a starburst galaxy should be $\sim 23$--33\% lower if one includes the nebular emissions in a model SED, consistent with the result of \citet{SdB09}. If we assume that this mass reduction also applies to IOK-1, the stellar mass upper limits would be slightly smaller, $M_* \leq 1.4$--$6.3 \times 10^8 M_{\odot}$ for the young, low dust scenario and $M_* \leq 0.7$--$2.8 \times 10^{10} M_{\odot}$ for the older or dusty scenario.

\section{Conclusion}
Using the {\it Spitzer} IRAC 3.6 and 4.5 $\mu$m imaging of the SDF, we study the stellar population of a $z=6.96$ Ly$\alpha$ emitter, IOK-1. The inspection of the images show that IOK-1 is not detected in these bands to $m_{3.6\mu{\rm m}} \sim 24.00$ and $m_{4.5\mu{\rm m}} \sim 23.54$ ($3\sigma$ limits). We fit population synthesis models to the SED of IOK-1 at optical to mid-infrared wavelengths to estimate the upper limit on its stellar mass. The IOK-1 can be either a low mass system with $M_* \lesssim 2$--$9 \times 10^{8} M_{\odot}$ for the young age ($\lesssim 10$ Myr) and the low dust reddening ($A_V \sim 0$) or a massive one with $M_* \lesssim 1$--$4 \times 10^{10} M_{\odot}$ for either the old age ($> 100$ Myr) or the high dust reddening ($A_V \sim 1.5$). The stellar populations of LAEs at $z\sim3$--6.6 have been studied to date and are known to include the galaxies from young, low mass, low dust ones (a few Myr, $\sim 10^6$--$10^8 M_{\odot}$, $A_V\sim0$--0.5) to old, massive, dusty ones (a few hundred Myr, $\sim 10^9$--$10^{10} M_{\odot}$, $A_V\gtrsim1$). Hence, the IOK-1 could be similar to one of these LAEs and is not an extremely high mass or low mass system.

The result presented in the current study are based on one LAE at $z\simeq7$, IOK-1. Thus, it might not precisely represent properties of the overall LAE population at $z\sim7$. Improving the statistics of $z \sim7$ LAE sample with deeper imaging and spectroscopic observations will help constrain the physical properties of $z\sim7$ LAE population more accurately. Also, a larger sample would allow stacking of the photometry to obtain a much deeper image and to study the average properties of $z\sim7$ LAEs. 

\bigskip

This work is based in part on observations made with the {\it Spitzer Space Telescope}, which is operated by the Jet Propulsion Laboratory, California Institute of Technology under a contract with NASA. Also, this work is based in part on data collected at Subaru Telescope, which is operated by National Astronomical Observatory of Japan (NAOJ). Use of the UKIRT 3.8 m telescope for the observations is supported by NAOJ. The WFCM data were reduced on the general common-use computer system at the Astronomy Data Centre (ADC) of NAOJ. We thank the referee for the useful comments that helped us to improve this paper. We thank Micol Bolzonella for the discussions about stellar mass estimate by using {\it Hyperz} . We express our gratitude to Tomonori Totani and Naoki Yasuda for providing us with $z'$-band imaging data to deepen the final stacked $z'$-band image, which is used for this work. We also thank Masami Ouchi for providing us with the $y$-band transmission data and the total $y$-band magnitude of IOK-1. We are grateful to Yosuke Minowa, Yoshiaki Ono, Toru Misawa and Poshak Gandhi for their advice and discussions about the data analyses and SED fitting. This research was supported by the Grant-in-Aid for Scientific Research (19540238) from the Japan Society for the Promotion of Science. K.O. acknowledges the fellowship support from the Special Postdoctoral Researchers Program at RIKEN where he completed most of the present work. T.M. is financially supported by the Japan Society for the Promotion of Science (JSPS) through the JSPS Research Fellowship. 



%
%


\end{document}